# H-Mode Accelerating Structures with PMQ Beam Focusing

S.S. Kurennoy, L.J. Rybarcyk, J.F. O'Hara, E.R. Olivas, and T.P. Wangler

*Los Alamos National Laboratory, Los Alamos, NM 87545, USA*

**Abstract.** We have developed high-efficiency normal-conducting RF accelerating structures by combining H-mode resonator cavities and a transverse beam focusing by permanent-magnet quadrupoles (PMQ), for beam velocities in the range of a few percent of the speed of light. The shunt impedance of inter-digital H-mode (IH-PMQ) structures is 10-20 times higher than that of a conventional drift-tube linac, while the transverse size is 4-5 times smaller. Results of the combined 3-D modeling – electromagnetic computations, multi-particle beam-dynamics simulations with high currents, and thermal-stress analysis – for an IH-PMQ accelerator tank are presented. The accelerating field profile in the tank is tuned to provide the best propagation of a 50-mA deuteron beam using coupled iterations of electromagnetic and beam-dynamics modeling. Measurements of a cold model of the IH-PMQ tank show a good agreement with the calculations. H-PMQ accelerating structures following a short RFQ can be used both in the front end of ion linacs or in stand-alone applications.

## I. INTRODUCTION

H-mode cavities (also called TE-mode cavities) were suggested for ion acceleration by Wideröe in 1928 [1] but since then have rarely been used in accelerators. One noticeable exception is heavy-ion accelerators where low-frequency RF (typically, tens of MHz) is required. For such low-frequency machines the transverse size of conventional drift-tube linac (DTL) structures would be enormous, of the order of a few meters, while with H-resonators it is a few times smaller. One important reason preventing H-cavity implementation in accelerators was the fact that their detailed design requires three-dimensional (3-D) electromagnetic calculations or, alternatively, an extensive experimental modeling. On the other hand, the well-known DTL structures [2] can be designed essentially analytically; having a simple 2-D eigensolver like Superfish [3] practically eliminates the need for multiple models. The DTL accelerators achieve their best efficiency for particle velocities $v$ in the range approximately from 10% to 35% of the speed of light $c$, i.e. $\beta = v/c = 0.1 - 0.35$. In the front end of the linear accelerator at the Los Alamos Neutron Science Center (LANSCE), the DTL is used in a wider velocity range, from $\beta = 0.04$ to 0.43, with a decreasing efficiency at the both ends.

One more H-mode structure, the radio-frequency quadrupole (RFQ) accelerator, was invented in the late 60s [4] and has received wide acceptance since the early 80s. The RFQ utilizes the quadrupole magnetic mode, $TE_{210}$, and provides not only acceleration but also effective electric focusing and beam bunching for very low-energy ions, typically at $\beta < 0.08$ [5]. These features make the RFQ an indispensible part of all modern high-current ion accelerators. On the other hand, it is well known that at its higher-energy end, $\beta > 0.03$, the RFQ is not a very efficient accelerator. The figure of merit for the accelerator efficiency, the effective shunt impedance $Z_{\text{eff}} = Z_{\text{sh}}T^2$ – e.g., see [5] – decreases as $\beta^{-2}$ along the RFQ length [6].

There is a gap in the accelerating efficiency between the two widely used structures, the RFQ and DTL, in the range of the beam velocities from $\beta = 0.03$ to 0.1. The H-mode room-temperature (RT) structures – inter-digital H (IH) that works in the dipole magnetic mode, $TE_{110}$, or the cross-bar H (CH) that employs the same quadrupole magnetic mode, $TE_{210}$, as the RFQ – can very well bridge this gap. This point is illustrated in Fig. 11 of a detailed review of H-type structures [7].

First, however, one has to address the beam transverse focusing. This is an important problem for all accelerating structures but especially those working at low beam velocities and with high beam currents. There are various options for the transverse focusing in linacs. In the RFQ the beam is focused by electric quadrupole RF fields created by four specially-shaped vanes or rods [5, 6]. This electric beam focusing is particularly effective at very low beam velocities since the magnetic Lorentz force – proportional to the velocity – is weak. In the DTL the beam focusing is produced by magnetic quadrupoles placed inside its large drift tubes (DT). In heavy-ion H-structures, the beam focusing is often provided by magnetic quadrupoles arranged in special focusing sections that interrupt the accelerating structure inside the tank, see in [7]. The focusing insertions increase the length and reduce the effective accelerating gradient. Such structures also use unconventional KONUS beam dynamics [8]. Another option – the only one available in superconducting cavities – is to provide magnetic focusing outside the tanks, between them. This leads to long distances between focusing elements, which is undesirable for low-velocity high-current beams. Finally, there is the alternating phase focusing (APF) that was recently implemented in the ion IH linac for hadron therapy in Japan [9]. For APF, the synchronous phases of the subsequent accelerating gaps are varied in a very wide range, from -90° to 90°, which reduces the structure accelerating efficiency. In addition, the APF scheme is usually limited to low currents.

We proposed [10] inserting permanent-magnet quadrupoles inside the small DT of the H-cavities; such structures can be called H-PMQ. This approach provides the transverse beam focusing in H-mode structures without any reduction of their accelerating efficiency. In particular, room-temperature IH-PMQ accelerating structures can bridge the efficiency gap between the RFQ and DTL, at the beam velocities in the range of a few percent of the speed of light.

In this paper we report results of developing H-PMQ structures, concentrating mostly on a particular example of a compact deuteron-beam accelerator up to the energy of a few MeV. Active-interrogation applications in homeland defense require deuteron beams of energy 4 MeV with the peak current of 50 mA and duty factor of 10%. One option to deliver such beams is a 4-MeV radio-frequency quadrupole (RFQ) accelerator. We suggest using IH-PMQ structures following a short RFQ as a more efficient solution.

The paper is organized as follows. In Section II we compare parameters of various low-energy room-temperature accelerating structures and present detailed characteristics of the H-PMQ structures. Section III is devoted to the design of a complete IH-PMQ tank, including the results of multi-particle beam-dynamics studies. In Section IV our modeling results are compared with measurements of the IH-tank cold model. Section V is devoted to discussion and conclusions.

## II. H-MODE ACCELERATING STRUCTURES

### H-mode Structures versus DTL

Let us compare a few low-energy accelerating structures, mainly focusing on their efficiency. The accelerator compactness, ease of fabrication, and overall cost are also important. Here we restrict ourselves to room-temperature (RT) structures only, to assure the system mobility and ease of use.

The radio-frequency quadrupole (RFQ) accelerator remains the best structure immediately after an ion source for accelerating light-ion beams with considerable currents. An RFQ is required to bring the deuteron beam to about 1 MeV while providing its bunching and transverse focusing. Assuming that, we consider alternative structures for the beam velocity range of $\beta = 0.034 - 0.065$, which corresponds to the deuteron kinetic energy from 1 MeV to 4 MeV. We assume the RF frequency around 200 MHz, which is in the range of both 4-rod and 4-vane RFQ designs. High RF losses exclude RT quarter-wave and half-wave resonators which are very efficient in low-energy superconducting (SC) accelerators for heavy ions [6]. Remaining options include the venerable drift-tube linac (DTL) and H-mode structures: IH (Inter-digital H) and CH (Cross-bar H, topologically similar to the spoke cavities). Table 1 lists their typical parameters in the low beam-velocity range; the data are compiled mainly from [5-7]. For RFQ, the shunt impedance is usually not cited; simple estimates give 2.6 MΩ/m for the SNS RFQ operating at 402.5 MHz.

**TABLE 1. Parameters of low-energy room-temperature accelerating structures.**

| Structure | "Best" $\beta$ | Frequency, MHz | $Z_{sh}T^2$, MΩ/m | Comments |
|---|---|---|---|---|
| RFQ | 0.005-0.03 | 4-rod: 10-200<br>4-vane: 100-425 | $\simeq 1-3$ | Estimate; $Z_{sh} \propto \beta^{-2}$; $T = \pi/4$ [5] |
| IH | 0.01-0.1 | 30-250 | $300 \rightarrow 150$ | $Z_{sh}T^2$ decreases as $\beta$ increases |
| CH | 0.1-0.4 | 150-800 | $150 \rightarrow 80$ | $Z_{sh}T^2$ decreases as $\beta$ increases |
| DTL | 0.1-0.35 | 100-500 | 25-50 | |

As a second reference point, we can take the average value of the effective shunt impedance $Z_{sh}T^2$ – the usual figure of merit for the accelerator efficiency – for the first tank in the LANSCE DTL ($\beta = 0.04 - 0.105$) at Los Alamos, 28.6 MΩ/m [5]. Based on these data one can expect that IH structures are an order of magnitude more efficient than DTL and RFQ at low beam velocities, for $\beta = 0.03 - 0.1$.

For a detailed comparison of the H-mode and DTL structures at the low-energy end ($\beta = 0.034$) of the deuteron accelerator, we calculated their parameters with the CST MicroWave Studio (MWS) [11]. The MWS eigensolver finds the modes in one structure period with periodic boundary conditions at the ends. One period contains two DT for H-cavities and one for DTL, as illustrated in Fig. 1, which shows one-period slices of four different accelerating structures. All the structures in Fig. 1 operate at the same RF frequency 201.25 MHz and have the same period, $L = \beta\lambda = 5.04$ cm, where $\lambda$ is the RF wavelength. We choose the gap length $g = 0.15 L_c$, where the cell length $L_c = \beta\lambda/2$ for H-cavities and $\beta\lambda$ for DTL. The DT bore (aperture) radius is chosen to be 0.5 cm; the outer radius is 1.1 cm for all H-types and 2 cm for DTL. The results are compared in Tab. 2, where field-dependent values are calculated for the average on-axis field $E_0 = 2.5$ MV/m; $R$ is the cavity inner radius; the maximal surface power density $(dP/ds)_{max}$ and power loss per period $P_{loss}$ are given at 100% duty for a copper surface with conductivity $\sigma = 5.8 \cdot 10^7$ S/m. The lengths of all cavity slices in Fig. 1 are the same but the transverse sizes are different, see $R$ in Tab. 2. The transit-time factors $T$ are close to 0.9 for all H-types in Tab. 2 and equal to 0.816 for the DTL. Note that the radius of the DTL is 4-5 times larger than in the H cavities. Moreover, the H-mode structures distribute the surface currents more evenly. All that reduces the power losses and makes the H-cavities much more efficient compared to DTL at these velocities. The vanes in H-cavities reduce the area of the regions of high power loss density on stems – cf. Figs. 1 (a) and (b) – and increase the structure efficiency even more.

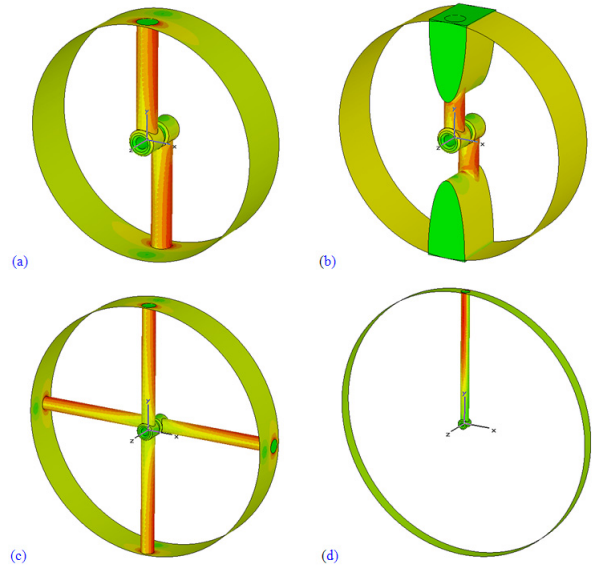

**FIGURE 1.** Surface current distributions for the structures in Tab. 2: (a) IH; (b) IH with vanes; (c) CH; and (d) DTL. Red means high current density, green – low. The spatial and current scales are different in all pictures, cf. Tab. 2.

Another explanation of the high efficiency of H-mode cavities follows from the field pattern. The IH magnetic field is directed along the cavity pointing in one direction on one side of the stems (one-half of the transverse cross section), and in the opposite direction on the other. The longitudinal electric field exists only because of the stems: the induced surface currents at a given time flow along one stem from the wall to DT and continue along the next stem from its DT to the wall. The electric field in the gap between these two DT closes the circuit. It is clear from the stem-current pattern that the electric field in the next gap is in the opposite direction. As a result, the electric field in H-cavities is concentrated mainly in small regions near the gaps.

**TABLE 2.** Structure comparison at $\beta = 0.034$.

| Structure | $R$, cm | $Z_{sh}T^2$, MΩ/m | $(dP/ds)_{max}$, W/cm² | $P_{loss}$, kW | $E_0TL$, kV |
|---|---|---|---|---|---|
| IH | 9.9 | 294 | 7.30 | 0.87 | 113.4 |
| IH w/vanes | 10.4 | 346 | 5.88 | 0.74 | 113.4 |
| CH | 16.4 | 227 | 4.60 | 1.13 | 113.4 |
| DTL | 55 | 21.5 | 31.1 | 9.74 | 102.0 |

For CH cavities, the picture looks similar except that the dipole pattern in the transverse cross section is replaced by a quadrupole one: the magnetic field points along the CH cavity in one direction in two diagonally-opposing quadrants, and in the opposite direction in two other quadrants. The surface currents flow from the wall to DT along one bar (a pair of stems connected to one DT), and from DT to the wall along the next bar (perpendicular to the first one), creating the longitudinal electric field in the gap between the two DT.

At the high-energy end of our deuteron linac, at $\beta = 0.065$, the differences in efficiency and size between H-structures and DTL become somewhat smaller but still are very significant, see [10] for details. One should mention that the accelerating structures presented in Tab. 2 and Fig. 1 are not optimized. Adjusting their parameters like drift tube, stem, or vane sizes can further improve their characteristics. Important to emphasize that increasing the outer diameter of DT in H-mode structures reduces the shunt impedance significantly; the DTL, however, is less sensitive to such a change. On the other hand, the H-structures are less sensitive to the change of the aperture size than DTL. An example of how the IH-structure efficiency can be increased by changing the vane and stem shape and the DT transverse sizes is given in [10]: a modified IH structure with deep vanes and small DT radii has the shunt impedance of 746 MΩ/m, more than double the best value in Tab. 2. For a fair comparison, the DT outer radius becomes only 0.75 cm and the bore radius 0.3 cm in that case. It would be very difficult to provide the transverse beam focusing in this structure if a strong PMQ needs to be placed inside such a tiny DT. One has to study the beam dynamics in the structure within practical limitations imposed by the existing PMQs and the requirements of the structure cooling.

## PMQ Focusing in IH Structures

To evaluate feasibility of PMQ focusing in IH-PMQ structures, we first performed beam-dynamics modeling with envelope codes code TRACE-3D [12] and its GUI version PBO-Lab [13]. To keep the DT length $L_{DT} = L_c - g$ as long as possible, we consider first the IH structure with narrow gaps $g$ between DTs by fixing the ratio $g / L_c = 0.15$. A 2-cm long PMQ with the bore radius 5 mm can readily provide the transverse magnetic field gradient $G = 200$ T/m, even for the PMQ outer radius of only 11 mm. It can fit into a DT even at the lower end of the IH accelerating structure, with geometrical value $\beta_g = 0.034$, where $L_{DT} = 2.16$ cm. We have considered the transverse focusing structure F$n$OD$n$O, where the focusing period consists of one focusing (F) and one defocusing (D) PMQ separated and followed by $n$ empty DTs, $n = 0,1,2,…$ We denote such structures as IH 1-(n+1), with one PMQ per $(n+1)$ cells. Beam dynamics simulations were performed with the envelope codes for $\beta_g = 0.034$ and 0.065. All field-dependent results are calculated assuming the average on-axis field $E_0 = 2.5$ MV/m; the RF synchronous phase in the gaps is -30°. The results are illustrated in Fig. 2 for IH1-3 ($n = 2$), where a PMQ is inserted in every third DT; the envelopes for 50 mA (outer lines) and zero current are plotted. The rms normalized transverse emittance $0.2\pi$ mm·mrad corresponds to $5 \cdot 0.2 / (\beta\gamma) \simeq 30\pi$ mm·mrad for the un-normalized emittance of the TRACE 3-D equivalent uniform beam at $\beta = 0.034$.

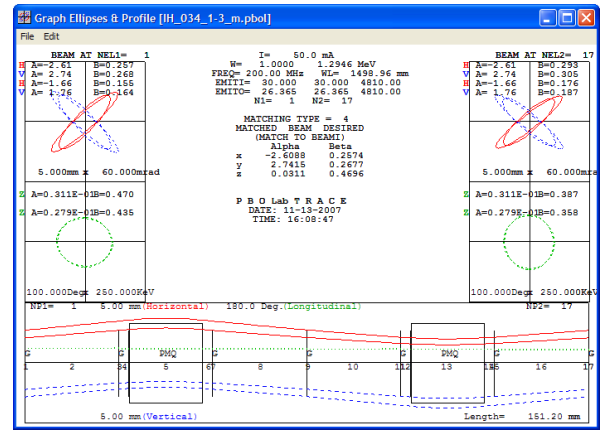

**FIGURE 2.** Beam matching results and envelope traces for one focusing period of IH1-3 structure at $\beta = 0.034$.

We calculated the beam sizes and phase advances per focusing period at both the low- and high-energy ends. At $\beta = 0.034$, for the IH1-1 case ($n = 0$, FD) 200-T/m quadrupoles are too weak to keep the beam size within the chosen aperture of radius 5 mm. Due to a very short distance, 0.52 cm, between the edges of F and D quads at the low-energy end – twice as small as the quad aperture – they noticeably cancel each other. Some long focusing periods, e.g. IH1-5, are excluded since their zero-current phase advances are above 90°, while the full current advances are below 90°; the beam can be unstable. All configurations IH1-2 to IH1-4 are acceptable, and the differences between them are not very significant; overall, IH 1-3 provides the smallest beam size. However, the beam size is rather large in all these cases, which can lead to undesirable beam losses. The transverse beam size variations along the period are obviously larger for IH1-4, especially compared to IH1-2 where they are minimal. The configuration IH1-2 requires placing a PMQ in every other DT; in IH1-4 the PMQs are placed only in every fourth DT, which provides a significant cost advantage.

Similar calculations were performed at the high-energy end of the linac, for $\beta_g = 0.065$. Due to longer periods, the zero-current phase advances are above 90° already for IH1-4. However, there are more options at the high-energy end compared to the low-energy end since the DT lengths are longer. We found that using longer PMQ while simultaneously increasing the PMQ and DT apertures to prevent beam losses gives better results. The modified IH1-3 structure has long PMQs, $L_q = 3$ cm, with weaker gradient, $G = 150$ T/m, which allows increasing the PMQ inner radius to $r_{in} = 6$ mm, with the outer radius $r_{out} = 12$ mm. The maximal beam size $r_{max} = 3.55$ mm is small compared to the 6-mm aperture radius, which is especially important to prevent beam losses at the high-energy end. More details and results can be found in [14, 15].

It is important to mention that the focusing strength of PMQ magnets strongly depend on the magnet radii, especially the inner one. For example, gradient $G$ of a typical PMQ with $M = 16$ trapezoidal permanent-magnet segments, the inner radius of $r_{in}$, and outer one $r_{out}$ is given [16] by

$$G = 2B_r \left(\frac{1}{r_{in}} - \frac{1}{r_{out}}\right) K_2, \qquad (2.1)$$

where $B_r$ is the remnant field, typically around 1 T, and the geometrical coefficient

$$K_2 = \cos^2 \frac{\pi}{M} \frac{\sin(2\pi/M)}{2\pi/M} \simeq 0.937. \qquad (2.2)$$

The simplified beam dynamics modeling with envelope codes indicates that IH-PMQ structures can work at low beam velocities for noticeable currents. We will follow with multi-particle beam dynamics results in Sec. III. So far, overall, the transverse focusing structure IH1-3 ($n = 2$, FOODOO), where PMQs are inserted only in every third DT, appears to be the best choice. It provides an acceptable beam transverse size while reducing the number of the required PMQs by a factor of three compared to the maximum equal to the number of DTs – a significant cost saving. It also gives us an opportunity to use DTs of different sizes – increasing the transverse size of DTs with PMQ while reducing the sizes of empty DT, as illustrated in Fig. 3 – to keep or even increase the high accelerating efficiency of the IH structure.

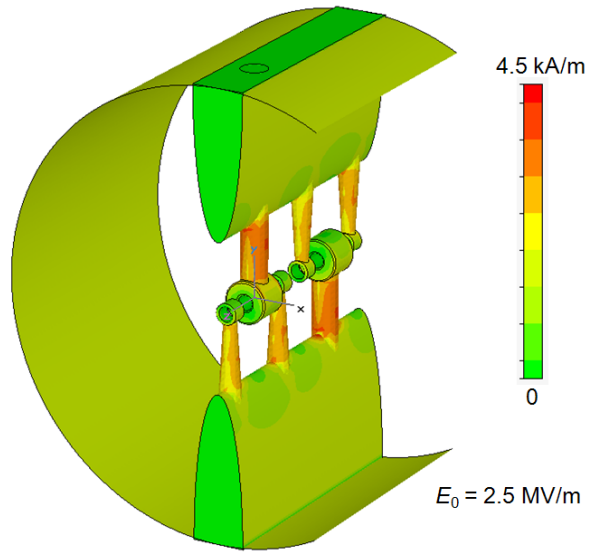

**FIGURE 3.** Surface-current-magnitude distribution in the modified IH1-3 structure for $\beta_g = 0.04$ (the cavity wall is partially cut for better view).

## IH-PMQ Linac: Cooling

One of the conditions for high efficiency in the H-PMQ structures is to keep the DT transverse sizes small. There is no space for cooling channels inside the small DT that already encloses a PMQ. Therefore, we should prove that it is feasible to cool the H-PMQ structures with cooling channels located only in vanes or outside manifolds, not in DTs. First let us estimate some parameters of our 1-4-MeV deuteron linac. With the average on-axis field $E_0 = 2.5$ MV/m, transit-time factors $T \simeq 0.9$, and all RF phases in the gaps at -30°, the energy gain of 3 MeV is achieved in approximately 1.5 m. Using power-loss values for IH structures with vanes from Tab. 1, and similar values at $\beta_g = 0.065$ [14], one can estimate the total surface power loss as 25 kW at 100% duty. This is a rather small fraction of the power delivered by the linac to a 50-mA CW beam, 150 kW. It is also an order of magnitude lower than in an equivalent DTL, ~250 kW.

For an accurate thermal and stress analysis we calculate the distribution of power deposited on the cavity walls due to the resistive loss from surface currents with the MWS, cf. Fig. 1. We have developed a procedure [17] to transfer surface-loss power data calculated by MWS to the finite-element (FE) engineering codes COSMOS [18] and ANSYS [19]. The important feature of this procedure is that the MWS fields are extracted not exactly at the cavity surface points but with a small offset into the cavity along the normal to each surface FE (triangle) out of the center point of the FE. This helps avoiding errors in the surface fields introduced by the hexahedral MWS meshes as well as in the cases when the central points of the surface FEs are located below the metal boundary, inside convex metal walls.

Thermal and stress analysis was performed for the regular IH structures with cooling channels located inside the vanes. For nominal 10% duty factor, the temperature distributions calculated in [20] by ANSYS are shown in Fig. 4.

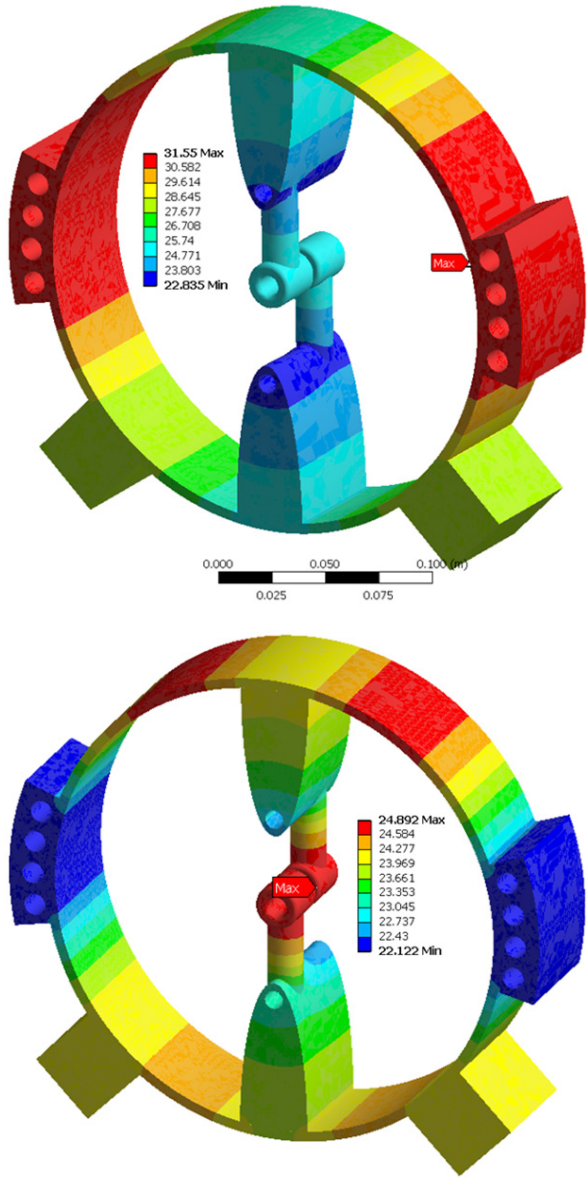

**FIGURE 4.** Temperature distribution in one period of regular IH structure with water cooling in vanes only (top), and both in vanes and outer manifolds (bottom).

In the top picture, the water cooling is applied only in the vanes: one channel near the top of each vane carries a 4.4-m/s water flow at 22°C inlet temperature. The maximal temperature (red) is 31.6°C, while the minimal (blue) one is 22.8°C. When the side-wall manifold cooling added at 10% duty (Fig. 4, bottom), the temperature range is only from 22.1°C to 24.9°C. Such an option may be needed at higher duty factors. This important result – the PMQ temperatures can be kept low by cooling only the vanes – confirms the feasibility of the H-PMQ room-temperature concept.

The DT relative vertical displacements for 10% duty are between 15 and 35 μm depending on the cooling scheme, below typical manufacturing tolerances. The stresses at 10% duty are practically the same as in the cold state due to the atmospheric pressure, below 8 MPa, and even at 100% duty they do not exceed 28 MPa, which is still very far from the copper yield stress of 57 MPa. The transient thermal-stress analysis did not show any stresses above the static ones.

## IH Parameters versus Beam Velocity

Before proceeding with the IH-PMQ linac design, let us consider how the characteristics of IH-PMQ structures depend on the beam velocity in the velocity range of interest $\beta = 0.034 - 0.065$. We take the IH structure with vanes shown in Fig. 1(b), operating at the RF frequency of 201.25 MHz, and having the fixed ratio of the gap $g$ length to the cell length $L_c$, namely $g/L_c = 0.15$. The vane height is 2/3 of the cavity radius, the DT aperture radius is 0.5 cm, and the outer DT radius is 1.1 cm, the same as in Tab. 1. Increasing the cell length according to the synchronism relation $L_c = \beta_g \lambda / 2$ and adjusting the frequency by small changes of the cavity radius $R$, we calculate the structure parameters with MWS for various values of the design velocity $\beta_g$. The results are listed in Tab. 3, where all field-dependent quantities are scaled to the average on-axis field $E_0 = 2.5$ MV/m, and power-related values are given at 100% duty. One can notice that the effective shunt impedance starts to decrease as velocity increases, which is the known property of IH structures [7]. However, it still remains an order of magnitude higher than the DTL shunt impedance in this velocity range.

The dependence of the transit-time factor $T(\beta_g, \beta)$ on beam velocity $\beta$ is plotted in Fig. 5 [14, 15] for a few structures with different design velocities $\beta_g$. The usual transit-time factors – defined by $T = T(\beta_g, \beta_g)$ and indicated in Fig. 5 by small red circles – increase slightly when $\beta_g$ increases, cf. Tab. 3.

**TABLE 3.** Characteristics of IH structures with vanes versus the design beam velocity

| Design beam velocity $\beta_g$ | 0.034 | 0.04 | 0.05 | 0.06 | 0.065 |
|---|---|---|---|---|---|
| Cell length $L_c$, cm | 2.52 | 2.98 | 3.72 | 4.47 | 4.82 |
| Transit-time factor $T$ | 0.900 | 0.915 | 0.939 | 0.953 | 0.958 |
| Quality factor $Q$ | 11780 | 12931 | 14012 | 14773 | 15313 |
| Effective shunt impedance $Z_{sh}T^2$, MΩ/m | 347 | 359 | 334 | 301 | 294 |
| Maximal electric field $E_{max}$, MV/m | 24.9 | 25.1 | 28.5 | 30.8 | 31.8 |
| Maximal surface-loss density $(dP/dS)_{max}$, W/cm² | 6.2 | 8.5 | 13.0 | 18.8 | 21.8 |
| Surface-loss power $P_{loss}$, W/cell | 368 | 435 | 614 | 842 | 940 |
| Effective voltage $E_0TL$, kV/cell | 56.7 | 68.2 | 87.4 | 106.4 | 115.4 |

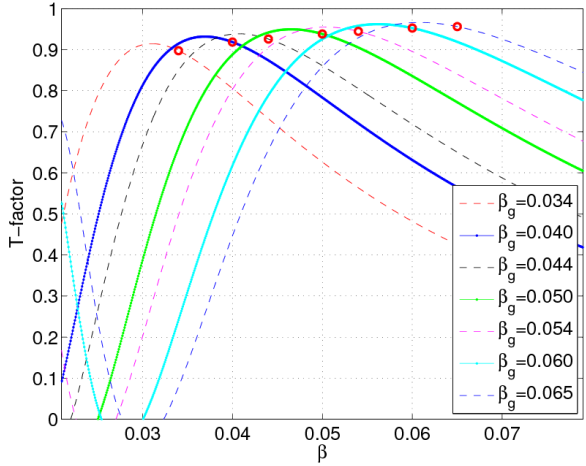

**FIGURE 5.** Transit-time factors of a few IH structures with vanes (defined by the cavity design velocity $\beta_g$) versus the beam velocity $\beta$.

Table 3 shows that the maximal surface electric field $E_{max}$ increases for higher design velocities. The values of maximal electric field $E_{max} \leq 1.8 E_K$, or even up to $2E_K$, where $E_K$ denotes the Kilpatrick field [5] at a given frequency, are considered safe for high-duty operations. For 201.25 MHz the Kilpatrick field $E_K = 14.8$ MV/m, so the maximal electric field in Tab. 3 exceeds $2E_K$ for $\beta_g \geq 0.06$. Since the electric field is concentrated in the gaps, one obvious solution for reducing the maximal electric field at a fixed gradient in the IH structures is to increase the gap lengths between the drift tubes by making the DTs shorter. This can be an attractive option at $\beta_g \geq 0.05$, since the drift tubes are relatively long so that they can still accommodate PMQs even when the DT lengths are reduced.

The data in Tab. 3 allow us evaluating possible designs of a 1-4 MeV deuteron accelerator based on the regular IH structures with vanes [14, 15]. We found only small differences between various designs, from one with gradually increasing cell lengths to the three-step design that includes only cells $\beta_g = 0.04$, 0.05, 0.06. In all these cases, the accelerator with a gradient of 2.5 MV/m and RF synchronous phase -30° throughout consists of 19-20 IH periods (38-40 cells) and has the total length 1.45-1.5 m. The surface-loss power is about 25 kW at 100% duty, which is small compared to the beam power of 150 kW for the 50-mA CW beam current. One remaining concern for the considered IH structures with narrow gaps is high values of the maximal electric field at higher beam velocities.

## Improving IH Structure Characteristics

Let us explore how the changes in the IH structure design such as wider gaps or smaller transverse sizes of the DTs influence the structure characteristics. The gap increase is achieved by reducing the DT length, while keeping the cell length fixed. Reducing the DT transverse size is another option for improving the characteristics of the H-mode resonators [7]. Both paths have obvious limitations in the H-PMQ structures where DTs have to house PMQs. Here we restrict ourselves to the case of the IH cavities with vanes discussed above.

Increasing the gap width by reducing the DT length, with other dimensions fixed except for a small change in the cavity radius to adjust the frequency, indeed helps. Apart from a small drop in the T-factor value (3-8%), all other parameters, and especially the shunt impedance, improve significantly when the gap increases from 15% to 30-45% of the cell length. The results of MWS calculations for IH structures with increased gaps [14] at $\beta_g$ = 0.04, 0.05, and 0.06 are plotted in Figs. 6 and 7. The maximal electric field is reduced to safe levels in the structures with larger gaps, see Fig. 6. The shunt impedance becomes higher, as illustrated in Fig. 7.

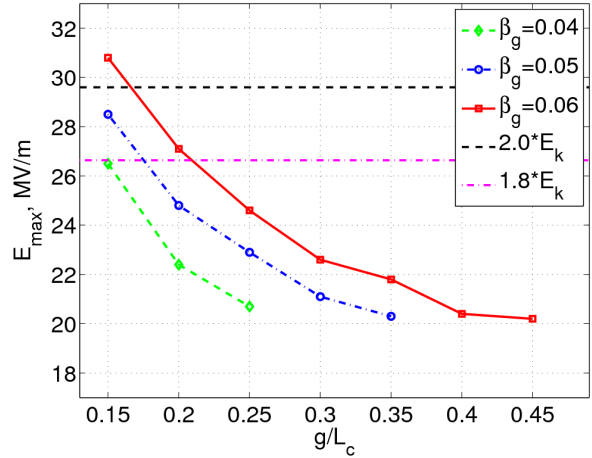

**FIGURE 6.** Maximal electric field in the IH structures with vanes versus the relative width of the gap between DTs.

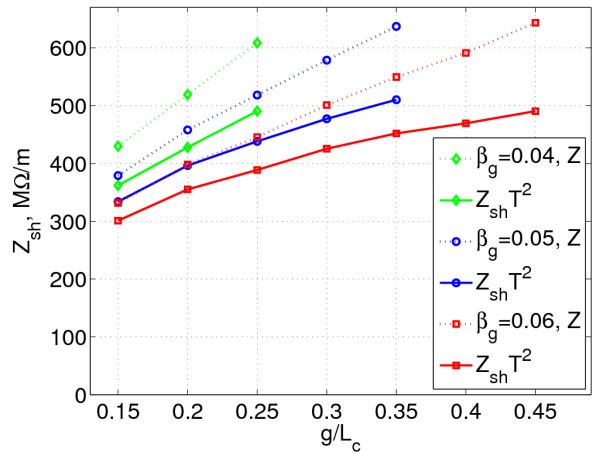

**FIGURE 7.** Shunt impedance (dotted lines) and effective impedance (solid) of the IH structures with vanes versus the relative width of the gap between DTs.

Another option to improve the IH-PMQ structure is to use DTs of different transverse sizes depending on whether they house PMQ inside or not. In IH1-3 structure discussed above, the transverse size of the DT with PMQ can be increased to facilitate the PMQ placement inside it, while the outer diameter of empty DTs can be reduced to keep the shunt impedance high. One can go a step further and reduce also the lengths of empty DTs to have wider gaps. One period of the IH1-3 structure modified in such a way was shown in Fig. 3. There the DTs with PMQ have large $r_{out}$ = 14 mm and length 24 mm; the empty DTs are short and slim, $r_{out}$ = 7 mm and length near 12 mm; the aperture radius is kept at 5 mm throughout. The resulting effective shunt impedance $Z_{sh}T^2$ is 712 MΩ/m for the structure of Fig. 3 with $\beta_g$ = 0.04. Even at the high-energy end of our small accelerator, at $\beta_g$ = 0.06, it is still above 500 MΩ/m [14, 15].

One should note that in IH resonators with wide gaps the on-axis electric field acquires a noticeable transverse component, especially when the outer DT radius is just slightly larger than the inner one (slim DTs). This effect has been observed long time ago in the IH accelerators for heavy ions, see in review [7]; it is caused by the inherent vertical asymmetry of the IH structures. With wide gaps between the DTs, parts of DT stems contribute to formation of the electric field in the gaps that connects the surface currents from one stem to two adjacent ones, cf. Fig. 8 (top).

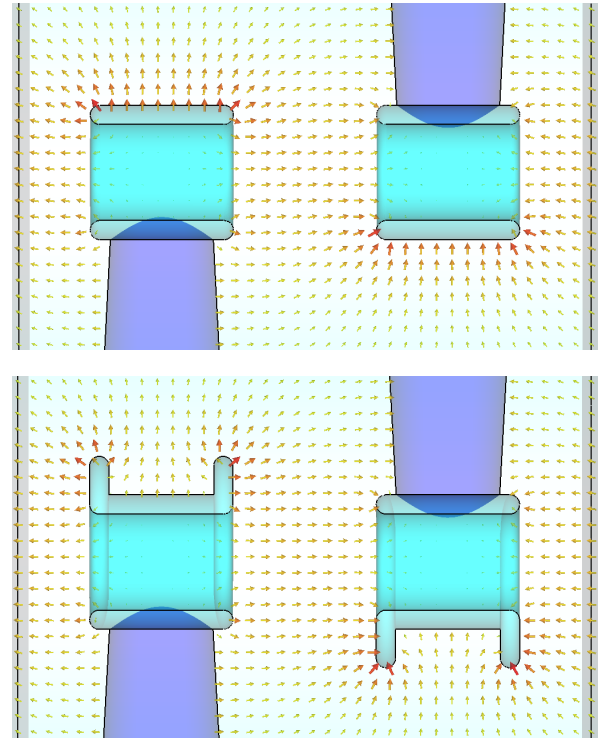

**FIGURE 8.** Electric field arrows in the axial vertical plane for one IH period ($\beta_g = 0.04$) with wide gaps, $g/L_c = 0.5$: standard slim DTs (top) and same DTs with bulges (bottom).

Asymmetric bulges on the DT outer surface, like ones shown in Fig. 8 (bottom), have been used to mitigate this effect. By introducing some compensating asymmetry in the DT shape, they reduce the transverse field in the gap. The longitudinal and vertical components of the electric field for one period of three IH structures are plotted in Fig. 9. All fields are normalized to $E_0 = 2.5$ MV/m.

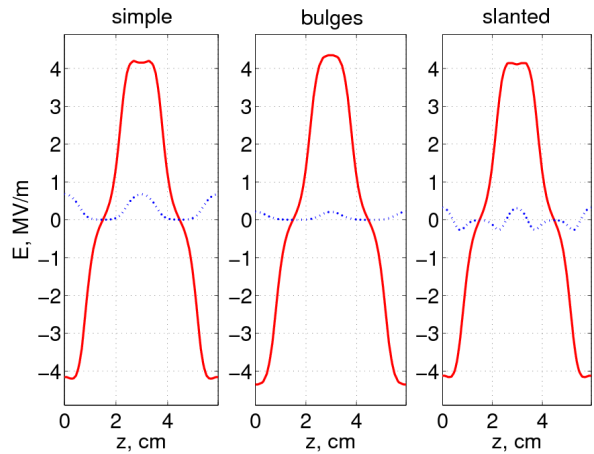

**FIGURE 9.** On-axis electric field components versus the longitudinal coordinate – longitudinal (solid red) and vertical (dotted blue): left and center – for the IH structures in Fig. 8, and right – for that in Fig. 10.

The bulges reduce the undesirable dipole field, but at the same time they reduce the shunt impedance: the structure with bulges in Fig. 8 (bottom) has the effective shunt impedance 20% lower than that without bulges, Fig. 8 (top) [14]. This effect is similar to that for DTs with a larger outer radius. In an alternative approach considered in [14], by making slanted the DT ends, see Fig. 10, the integrated transverse kick can be nullified with a proper choice of the slant angle, cf. Fig. 9 (right).

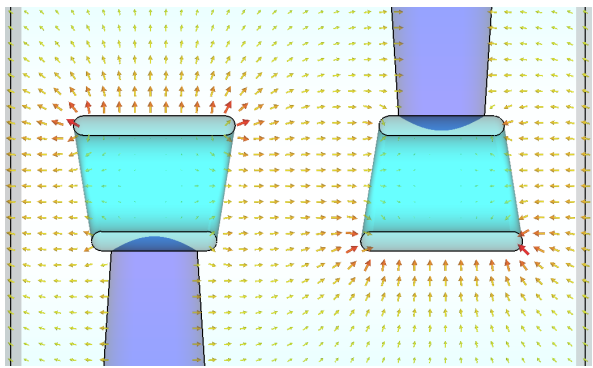

**FIGURE 10.** Electric field lines in the axial vertical plane of one IH period with wide gaps and DTs with slanted ends. The field logarithmic scaling is the same in Figs. 8 and 10.

The slanted-end DTs keep high shunt impedance but increase the maximal electric field slightly more than the bulges do. One can see in Fig. 9 that for wide gaps the accelerating-field profile in the gap becomes flat-topped (typically for $g/L_c = 0.45$-$0.5$) or even double-humped for $g/L_c > 0.5$ [14], instead of having a peak in the middle as in more narrow gaps. For even wider gaps, the transit-time factor quickly decreases. For example, in the IH1-3 structure of Fig. 3, the on-axis electric field has a large vertical component (~20% of the double-humped longitudinal one) in the wide gap between two slim short DTs.

We also notice in Fig. 9 that the period of the deflecting transverse field is twice as short as that of the accelerating field, so a beam being accelerated in the structure will experience alternating vertical kicks in subsequent gaps, which can compensate each other. However, effects of the on-axis transverse electric field on the beam in particular applications should be evaluated with multi-particle simulations.

In general, the IH-PMQ structures designed for high currents require rather strong transverse focusing, which can be achieved by placing PMQs in all DTs. The PMQ focusing strength is a product of its gradient $G$, Eq. (1.1), and the PMQ length $L_q$. This eliminates options of using slim or short DTs, if they are to house PMQs. One the other hand, since such structures will have relatively narrow gaps between DTs, $g/L_c \leq 0.3$, and DTs have some noticeable radial extent, there will be practically no on-axis transverse field in the gaps.

## III. IH-PMQ TANK DESIGN

We proceed now with a design of a resonator tank that contains an IH-PMQ structure. The magnetic field of the $TE_{110}$ mode, the working mode of IH resonators, is directed along the beam axis on one side of the DTs, their supporting stems, and vanes or girders if they are present. The field magnitude is practically constant in the resonator vertical transverse cross section [7]. The field returns on the other side of the DTs and their supporting elements. In other words, the magnetic field lines form a loop around the DTs and supports in the plane containing the beam axis and transverse to the stems. There should be some space between the vanes and the end walls of the resonator for the magnetic flux to turn around. The same applies to the higher-order H-modes, e.g., $TE_{210}$ used in RFQ and CH resonators. This consideration defines the design requirement for the tank end cells. Strictly speaking, one can talk about $TE_{n10}$ modes only in an infinitely long resonator: the third index equal to 0 means that the field is independent of longitudinal coordinate. The presence of end walls certainly violates this condition. This is why sometimes the working mode of an IH resonator is denoted $TE_{11(0)}$, meaning that along the large part of the resonator the field structure corresponds to the transverse waveguide mode $TE_{11}$.

Based on the above considerations, it would be reasonable to design our compact deuteron accelerator as a single resonator (tank). The cavity end walls introduce additional losses and reduce the structure efficiency so the number of tanks should be minimal within practical possibilities. However, one of our project goals was to fabricate and measure a cold model of the IH tank. To simplify this task, we decided to design a short tank containing the IH-PMQ accelerating structures with vanes. The range of the design beam velocities was chosen $\beta = 0.0325$-$0.05$, so that the tank IH1 can serve as the first of two tanks in our 1-4 MeV deuteron accelerator. We concentrate on the end-cell design and the means to tune the electric field profile along the beam axis, as well as the frequency of the working mode to 201.25 MHz.

## IH1 Tank Design

The tank IH1 layout [21] is presented in Figs. 11-12. Figure 11 shows only the IH DTs, supporting stems, and vanes. The vane elliptical transverse profile was changed to rectangular to simplify fabrication. The vane undercuts near the end walls allow for the magnetic-flux return. The cut dimensions are adjusted to reduce the electric field drop near the tank ends.

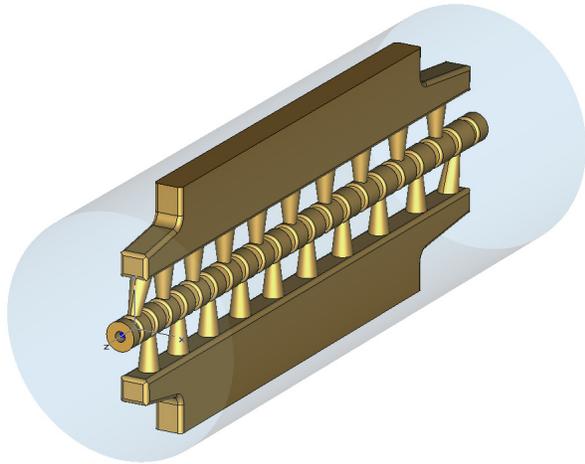

**FIGURE 11.** MWS model of the IH1 tank. The outer wall is removed; the cavity inner volume is shown in light-blue.

In Fig. 12 the longitudinal cross section of the tank is shown. This tank contains 10 IH "periods," each consisting of two cells. The total number of DTs is 20, and two more half-DTs are located on the end walls.

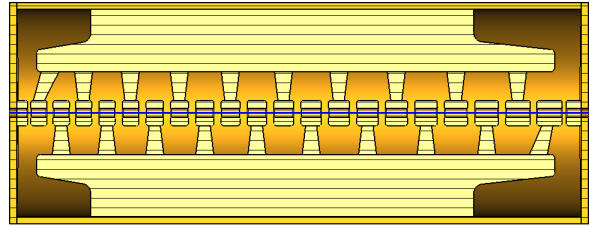

**FIGURE 12.** Longitudinal vertical cross section of the IH1 tank with the outer wall. The beam axis is shown by the blue line.

The cell length increases along the structure as $L_c = \beta_g \lambda / 2$ but most of the inner gaps have the same width of 8 mm. The widths of a few end gaps are reduced to 3.3, 6.8, and 6.8 mm in the upstream end (left in Fig. 12) and to 7.25 and 4.25 mm in the downstream end, to bring up the on-axis fields there. Some tank dimensions are listed in Tab. 4.

**TABLE 4. Dimensions of IH1 tank.**

| Dimension | Value, cm |
|---|---|
| Tank inner-cavity length | 62.97 |
| Inner radius of the tank cavity | 11.52 |
| DT length min / max | 1.81 / 2.84 |
| DT aperture radius / outer radius | 0.5 / 1.4 |

*IH1-Tank Electromagnetic (EM) Modeling*

EM modeling of the tank was performed with the CST MWS. The magnitude of the on-axis electric field of the main mode at 201.25 MHz in the IH1 tank is shown in Fig. 13 by a blue dash-dotted line, for the accelerating gradient $E_0 = 2.5$ MV/m.

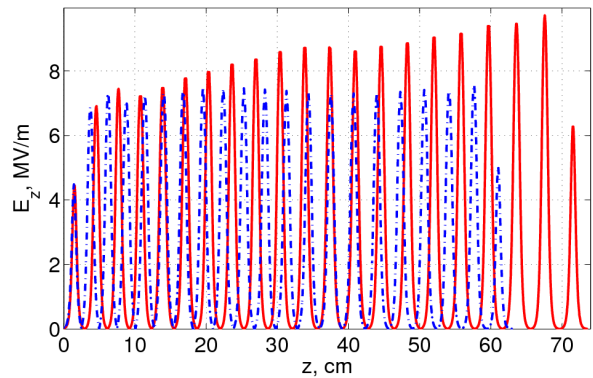

**FIGURE 13.** Longitudinal electric field profile along two tanks: IH1 – blue dash-dotted, and IH1m – red solid curve.

The next closest mode – $TE_{11(1)}$ longitudinal harmonic, with one node in the middle of the tank – has the frequency of 224 MHz, 23 MHz above the working mode.

The electric field profile of the working mode is relatively flat though the end-gap fields are lower than those in the inner gaps, in spite of the vane undercuts and reduced gap widths near the tank ends. The magnitude of the electric field in a gap is rather sensitive to variations of the gap width *g*: reducing the width of a single gap in the middle of the tank by 1 mm (12.5%) increases the local on-axis field by almost 10%, without large changes elsewhere. The frequency sensitivity to such a change is $\Delta f / \Delta g \simeq 0.4$ MHz/mm. The calculated electromagnetic parameters of the IH1 tank are summarized in Tab. 5. As before, all field-dependent quantities are scaled to the average on-axis field $E_0 = 2.5$ MV/m and power-related values are given at 100% duty for conductivity $\sigma = 5.8 \cdot 10^7$ S/m.

**TABLE 5. EM Parameters of IH1 tank.**

| Parameter | Value |
|---|---|
| Quality factor $Q$ | 9678 |
| Transit-time factors $T$ (21 gaps) | 0.88-0.94 |
| Effective shunt impedance, MΩ/m | 481 |
| Maximal electric field, MV/m | 20.2 (1.37$E_K$) |
| Surface RF power loss, kW | 13.4 |
| Maximal surface loss density, W/cm$^2$ | 69.2 |

The maximal value of the surface loss density is achieved on the edges of the vane undercuts as illustrated in Fig. 14. The loss density on the outer cover (not shown) and all inner DTs is relatively low. The loss power is distributed as follows: 44% on the outer cover, 36% on the vanes, 14% on the DT stems, and only 6% on the DTs. It is therefore important to provide adequate cooling of the tank vanes.

We have considered frequency and field-tilt slug tuners for the IH1 tank [21]. A symmetrical pair of cylindrical slugs of 8-cm diameter in the middle of the tank shifts the cavity frequency by $\Delta f / \Delta d \simeq 0.77$ MHz/cm, where *d* is the slug protrusion depth. The electric fields in the gaps near the tuners slightly decrease as the tuners move in. Two pairs of such slug tuners placed one near the beginning and one near the tank end can provide a tilted field profile if needed. Moving the first pair of slugs in and the second one out by 1 cm creates a tilt of about 16%, with the lower fields in the beginning of the tank. The opposite tilt (high to low with the slugs out / in) is less pronounced, about 10%.

*Thermal-Stress Analysis*

The heat loads on the tank surfaces were calculated using the MWS and transferred into ANSYS using the procedure [17] described above, for a thermal-stress analysis [21]. The tank is assumed to be water-cooled using one channel in each vane and one U-shaped loop on each of the two end walls. The temperature distribution for the duty factor of 10% is shown in Fig. 15. One can notice the PMQs (SmCo) in Fig. 15 shown as inserts inside the DTs; their material properties were taken into account in the thermal analysis. The inlet water is at 22°C, with the flow rate 2 gpm at velocity 4.5 m/s. For the 10% duty, the maximal temperature is 34.5°C on the first DT (red) while the minimal one is 23°C in the vanes (dark-blue). For higher duty factors, 15% and 20%, the maximum reaches 41°C and 47°C, respectively, for the same cooling, with the temperature pattern similar to that in Fig. 15. It is important to emphasize that the proposed cooling in vanes is sufficient to keep the temperatures of the PMQs inside the DTs well below the maximal acceptable PMQ working temperature, 150-250°C, without need to cool the DTs and stems using dedicated channels.

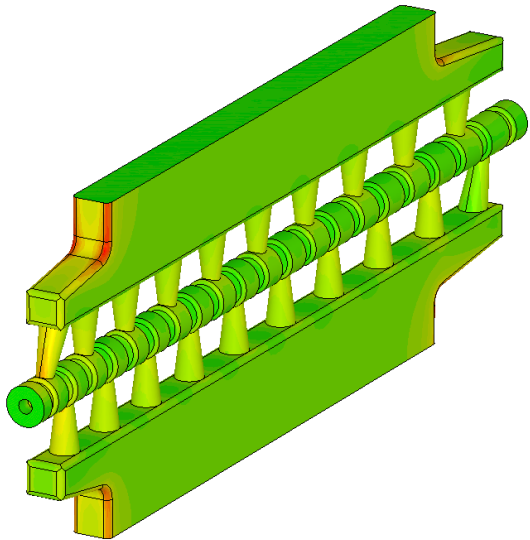

**FIGURE 14.** Distribution of surface-current magnitude in the IH1 tank: red color corresponds to the highest value, and dark-green to zero.

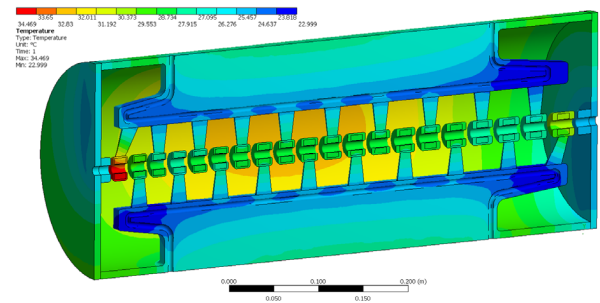

**FIGURE 15.** Temperature distribution in IH1 at 10% duty. Cut-out view reveals the cooling channels inside the vanes.

The maximal structural stress at the duty factor of 10% is only 37 MPa near the vane-cover connection and can be further reduced by making a smooth blend there. The relative longitudinal DT displacements in the structure are small, below a few μm. There are some vertical displacements of the DTs supported by the opposite vanes, especially near the tank ends: about 48 μm between the two first DTs (on the upstream end), and near 60 μm for the last two DTs. These misalignments can be prevented by choosing the initial positions of these DTs with proper vertical offsets, if beam dynamics simulations indicate that it is needed.

*Beam Dynamics Simulations for Tank IH1*

For the beam dynamics simulations, we assumed that each DT contains a PMQ. All PMQs are identical, with the inner radius 6 mm, outer radius 12 mm, and length 16 mm. The quadrupole gradient with SmCo was estimated to be 165 T/m. The field-overlap effect for the two close PMQs in the F-D configuration was calculated; it reduces the fields by less than 3% in the worst case. TRACE-3D envelope simulations [21] indicated that the best transverse focusing is achieved when the PMQs are arranged to form an FFDD lattice, see Fig. 16.

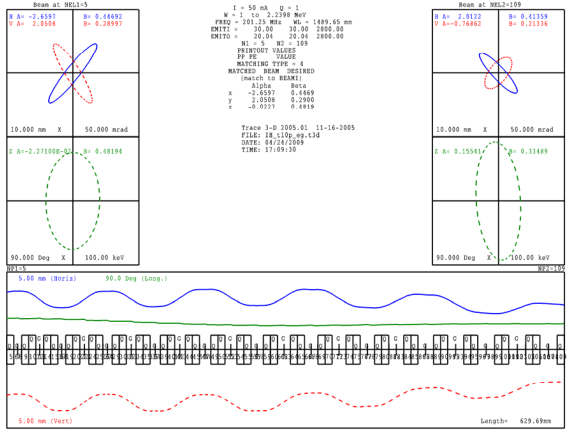

**FIGURE 16.** Output of TRACE-3D envelope calculations for the IH1 tank.

The average accelerating gradient per cell corresponding to the flat electric-field profile for the IH1 tank in Fig. 13 (blue dash-dotted) goes from high to low along the tank, due to the longer periods in its downstream part. Multi-particle beam dynamics simulations were performed for a 50-mA deuteron beam with the initial normalized transverse rms emittance of $0.2\,\pi$ mm·mrad at the source. First the beam propagation through a generic RFQ from 50 keV to 1 MeV was modeled with PARMTEQM [22]. The results were used to establish the parameters for the input beam emittance in a beam-envelope simulation of the IH tank using TRACE-3D. Matching the input beam in the first full IH period produced the beam envelopes shown in Fig. 16. The synchronous RF phase for each gap was -30°. However, multi-particle simulations performed with PARMILA [22] indicated particle losses on the DT walls.

We expected that a phase ramp in the upstream part of the tank would improve the beam dynamics. The phase ramp can be achieved by moving the DTs, while keeping the gap lengths fixed, in the longitudinal direction: to reduce the gap phase by 15°, the DT in the beginning of the tank should be moved upstream by less than 2.1 mm. The tank layout was modified so that the phase was ramped slowly from -45° to -30°, excluding the first and the last gap. This provided a smooth change in the longitudinal focusing along the structure while increasing the acceptance at the beginning. Since the first and last cells have $E_0$ that is about half of that for the neighboring cells, cf. Fig. 13, the synchronous phase of those cells was adjusted to set the RF-bucket height approximately equal to that of its neighbor, which resulted in a somewhat wider bucket as well. Still, when multi-particle simulations were performed for this design, about 5% of the input beam was lost. Even for a smaller initial rms emittance at the source, $0.13\,\pi$ mm·mrad, the particle loss was above 4%. Stronger space-charge forces for the smaller beam led to almost the same losses as for the larger beam. The main reason was that due to the small DT lengths in the IH1 tank, starting from 1.8 cm, the short PMQs inside the DTs were not strong enough to focus high-current beams.

## IH1m Tank Design

The IH1 tank was redesigned [23] to have higher injection deuteron energy of 1.5 MeV ($\beta_{in} = 0.04$), so that all DTs are longer in the new design and allow for longer PMQs. The higher injection energy also reduces the space-charge forces in the beam entering the tank. The output energy increased correspondingly to ~2.8 MeV, with $\beta_{out} = 0.0543$. The 3-D EM modeling of the modified tank, IH1m, was performed with the CST MWS as before, but this time we used iterations of beam-dynamics and EM simulations to adjust the tank layout. The gap widths were tuned for the electric field strength to increase along the tank proportionally to the cell length, cf. Fig. 13 (red solid curve), to keep the cell gradient nearly constant. The gap positions were also adjusted to create a ramp of the synchronous phase along the tank from -45° to -35°, except for the first and the last gap. For the first and last cells, which have lower cell gradients, the phases were adjusted to -

54° and -41°, respectively, to have the same RF-bucket height as the adjacent cells, similar to the IH1 tank. Such a ramp maintains a constant phase width of the separatrix and provides for better beam capture. The modified IH-PMQ tank is shown in Fig. 17. The cavity total length is 73.51 cm, and its radius is 11.92 cm.

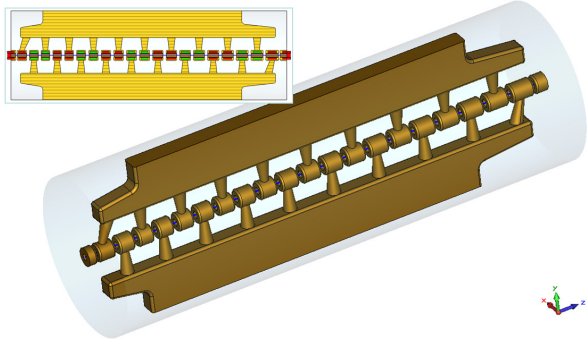

**FIGURE 17.** MWS model of the IH1m tank with the outer wall removed. The inset shows a longitudinal vertical cross section and PMQs inside the DTs.

The electromagnetic (EM) parameters of the tank for $E_0 = 2.5$ MV/m are listed in Tab. 6. Compared to the IH1 tank parameters in Tab. 5, the transit-time factors and Q are slightly higher but the effective shunt impedance is lower due to the higher design beam velocity, see Sec. II.

**TABLE 6. EM Parameters of IH1m tank.**

| Parameter | Value |
|---|---|
| Quality factor $Q$ | 9973 |
| Transit-time factors $T$ (21 gaps) | 0.91-0.95 |
| Effective shunt impedance, MΩ/m | 408 |
| Maximal electric field, MV/m | 23.2 (1.58$E_K$) |
| Surface RF power loss, kW | 19.2 |
| Maximal surface loss density, W/cm$^2$ | 113 |

In the IH1m tank the shortest DT is DT3 (23 mm); the longest one is DT20 (34.3 mm). The gap widths vary between 7.1 and 8.3 mm, except the very first and the last gap, which were reduced to 3.6 and 3.9 mm, respectively, to bring up the electric fields near the tank ends. Another change aimed at improving the beam dynamics is that all drift tubes up to DT11 have the bore radius of 5 mm, and from DT12 they have a larger bore radius of 5.5 mm. The outer radius of all DTs is the same, 14 mm. The field profile in the tank, as well as its frequency, can be tuned with two pairs of slug tuners in the side walls, one pair near the tank entrance and the other near its end. The slug tuners are not shown in Fig. 17 but can be seen below in Figs. 21-22 of the IH-PMQ tank cold model.

*PMQ Fields in IH1m Tank*

Two families of the 16-segment PMQs are used for beam focusing in this tank: the short ones, 18.89 mm, in the first 12 DTs (0 to 11), and longer ones, 22.67 mm, from DT12 on (12-21). The remanent magnetic flux density for the PMQ SmCo segments is 1.0 T, a conservative value. The inner PMQ radius is 5.5 mm for the first PMQ family, and 6 mm for the second; the outer radius is 11 mm for both. The quadrupole gradients of the 16-segment PMQs are 170 and 142 T/m, but the integrated focusing strengths for both PMQ types are the same, 3.2 T. The PMQ magnets are arranged in pairs to form an FFDD beam focusing lattice, as schematically illustrated in the inset of Fig. 17 (F in red, D in green).

To take into account the PMQ field overlaps, the static magnetic field for the whole array of 22 PMQs, shown in Fig. 18, was calculated by the CST Electro-Magnetic Studio (EMS) [11]. The computation was performed with a very fine mesh, 24M points for one quarter of the structure, with account of symmetry, but took only about 10 min on a PC with dual quad-core Intel W5590 3.33-GHz processors. The red segments in Fig. 18 were used in EMS computation, with the field symmetry taken into account. The field overlaps are clearly seen in the left inset of Fig. 18, which shows the vertical field component at the aperture of the smaller DTs in the horizontal plane ($x = 5$ mm, $y = 0$). The right inset shows B-field arrows in the middle of PMQ in DT10 (log scale, maximal arrow is 1.05 T).

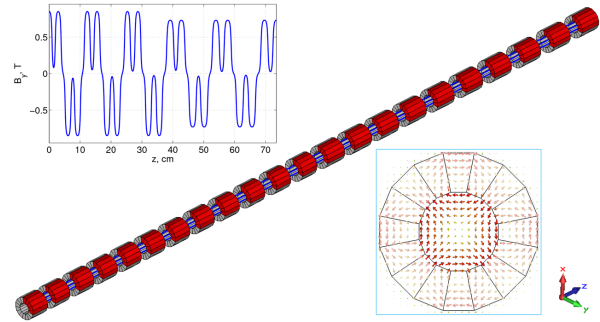

**FIGURE 18.** Array of 22 PMQs in the IH-PMQ tank and their magnetic fields calculated by CST EMS.

*Beam Dynamics Simulations for Tank IH1m*

As we already mentioned above, the design of the IH1m tank in Fig. 17 was optimized for high-current beams using iterations of 3-D EM MWS calculations and a specialized linac design code that we have developed based on the PARMILA [22] algorithms. The code applies the transit-time factors calculated for individual cells from MWS results to fine-tune the beam velocity profile along the tank. Multi-particle simulations with PARMILA did not show particle losses. After that we employed 3-D multi-particle beam dynamics simulations with PARMELA [22] and CST Particle Studio (PS) [11] to confirm the design. The emittances of the matched input beams to the IH tank were estimated using PARMTEQM [22] for a generic RFQ with the deuteron output energy 1.5 MeV and current 50 mA. The initial normalized transverse rms emittance in the RFQ was 0.13 π mm·mrad at 62 keV and 55 mA. This value is based on the beam emittance measured at the LEDA CW 100-mA proton source [24], and usual mass and current scaling arguments. Both PARMELA and PS multi-particle simulations have used the same exact 3-D fields from the CST Studio codes: RF fields from MWS and static magnetic field from EMS.

We explored both "water-bag" and Gaussian initial distributions of the bunch particles generated by the PARMILA code, with up to 100K particles used in simulations. The same initial distributions were imported into both PARMELA and PS runs. Results of the beam dynamics simulations for the IH tank indicate no particle loss at the fractional level down to $10^{-5}$ even at 50 mA, as illustrated in Figs. 19-20.

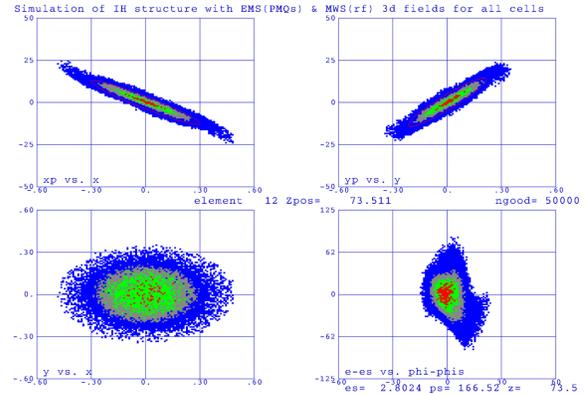

**FIGURE 19.** PARMELA phase-space plots of a Gaussian bunch with 50K particles at the IH1m tank exit.

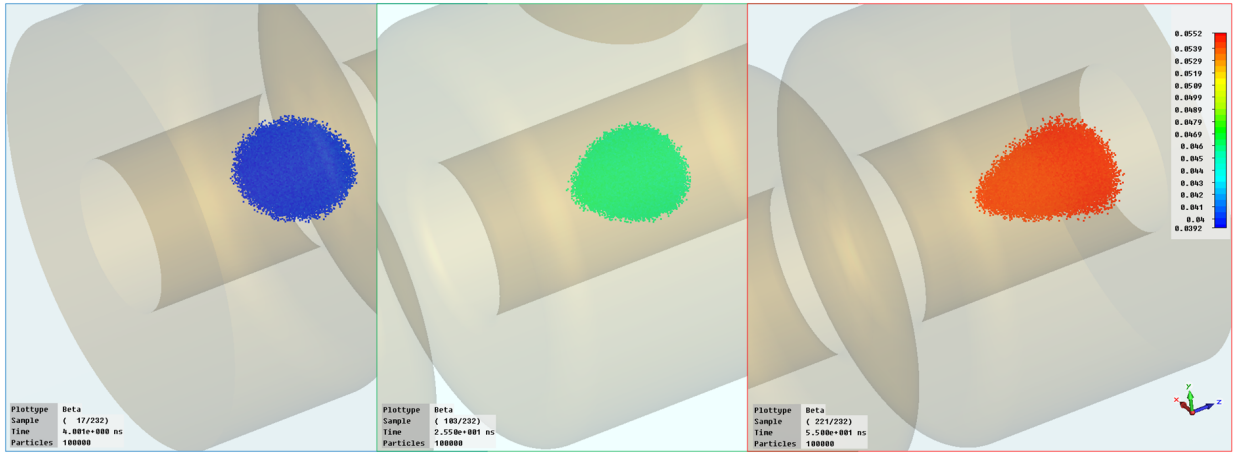

**FIGURE 20.** Evolution of a Gaussian bunch with 100K particles in the IH-PMQ tank computed by the CST Particle Studio: near the entrance, in the middle, and near the exit. Color changes indicate the particle-energy increase (β-scale shown).

The numbers for the emittance growth in the IH1m tank calculated by both codes are in agreement. The transverse normalized rms emittance increases 16-18% for the Gaussian initial distribution, and about 8% for the "water-bag" one. Such increases may look large, but we should recall that it is a high-current deuteron beam at very low beam velocities.

## IV. COLD MODEL OF IH-PMQ TANK

A cold model of the IH1m tank has been designed and manufactured. The cold model is a simplified structure made from aluminum for the purpose of making low-power field measurements; it does not have any active cooling features. Figure 21 shows an exploded view of the model. The outer shell of the tank is comprised of four longitudinal pieces as well as upstream and downstream end walls. Each individual drift tube was machined to finished dimension, and then, using a press-fit technique, was attached to its individual stem. The drift tube and stem assemblies were bolted to either the top or bottom vane pieces. The assembled vanes were then pinned and bolted to the top and bottom longitudinal pieces of the outer cover respectively.

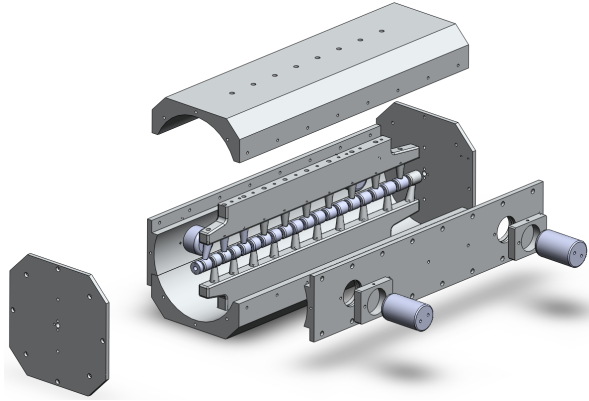

**FIGURE 21.** An exploded view of the tank cold model. Two pairs of slug tuners are located on the side panels.

In order to maintain the fidelity between the fabricated cold model and the computer simulations a significant amount of attention was spent on the alignment of the drift tubes. Precision Teflon spacers were fabricated and used to set the proper longitudinal gaps between drift tubes. The transverse alignment was set through the use of a precision alignment fixture that properly oriented the individual drift tubes to each other, and to the vane they were mounted on. Because of rather small gaps between the drift tubes, the cold model alignment is very important.

Figure 22 is a close-up photograph of the drift tubes in the bottom part of the model assembly. One can see open holes for the tuner slugs on the side wall behind the DTs.

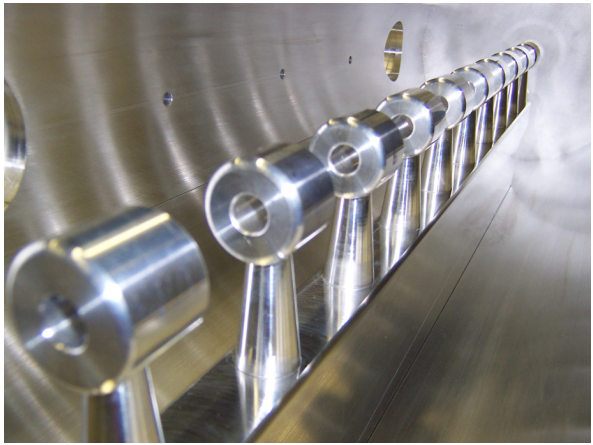

**FIGURE 22.** A close-up view of the DTs installed on the bottom part of the IH1m cold model.

For the cold model we measured the mode frequency and the electric field profile with a bead-pull technique. The measured field profile is in a good agreement with the MWS calculation. The comparison between the measured and calculated values of the cell gradient $E_{0c}$ is shown in Fig. 23, where both computed and measured values are scaled to the same tank gradient $E_0 = 2.5$ MV/m.

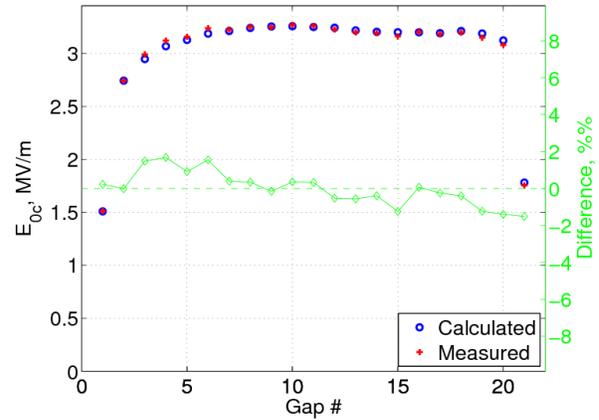

**FIGURE 23.** Calculated and measured cell gradients in the IH1m tank cold model (left vertical axis) and differences between them (right vertical axis).

The differences, shown on the right vertical axis, are within 2%, and indicate a slight tilt from upstream to downstream. A perfect agreement was not expected since the measurements were done with the slug tuners retracted from their nominal positions to bring the frequency down.

The measured frequency of the cold model with the slug tuners in the nominal position turned out about 1.35 MHz (0.67%) higher than the calculated design value of 201.25 MHz. Such a difference was rather unexpected since, from our previous experience, the frequencies computed by the MWS with reasonably dense meshes are quite accurate. The design frequency was predicted [23] based on the MWS computations using rather dense meshes up to 8-9 million points covering one half of the model; the computations take into account the tank symmetry with respect to its vertical axial plane.

After the discrepancy was found, we revisited our frequency computations. The MWS eigensolver runs were repeated with varying number of mesh points $N$, both with and without adapting meshing. We found a rather peculiar asymptotic behavior: the calculated frequency continued to monotonically increase very slowly as the number of mesh points increased, see in Fig. 24. Typically the calculated frequency changes first quickly as the number of mesh points $N$ increases, and after that, for sufficiently large $N$ values, its value wobbles in a narrow range.

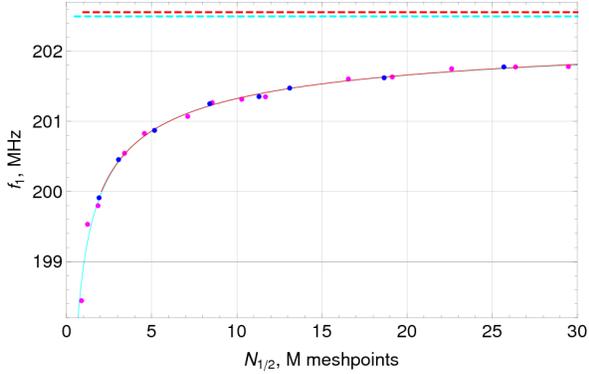

**FIGURE 24.** Fitting the calculated mode frequency of the cold model versus the number of mesh points used in MWS computations. Blue dots are results without adaptive meshes; they are fitted with the red solid curve; the dashed red line shows the corresponding asymptotic frequency. Magenta dots are for results with adaptive meshing. They are fitted by the cyan solid curve, and the cyan dashed line indicates the asymptotic frequency.

The computation results plotted in Fig. 24 suggest fitting the frequency dependence as

$$f(N) = f_o - bN^{-\alpha}, \qquad (4.1)$$

where $f_0$ is the asymptotic frequency to be found, $b$ and $\alpha$ are two positive constants. One can expect from physical considerations that $1/3 \leq \alpha \leq 1$, where the lower bound seems natural for cases with spherical symmetry. Since our model layout is approximately cylindrical, we expect the exponent to be closer to 1/2. We find the parameters of the fit (4.1) from the data in Fig. 24 using Mathematica [25]. Two sets of data give two similar sets of fit parameters: for meshes without adaptive refinement, $f_0 = 202.55$ MHz, $b = 3.554$, and $\alpha = 0.461$ (red curves); with adaptive meshes, $f_0 = 202.50$ MHz, $b = 3.557$, and $\alpha = 0.483$ (cyan curves). The fit curves in Fig. 24 practically overlap. The calculated asymptotic values of frequency are close to the measured value.

The MWS eigensolver run with the densest mesh of 29.5M mesh points (the right-most magenta point in Fig. 24) took ~31 hour on a PC with dual quad-core Intel W5590 3.33-GHz processors. One should add that we have also performed MWS computations using meshes with different mesh parameters (the ratio of the largest to the smallest cell size, etc). The fit curves looked different (the fit parameters $b$ and $\alpha$ take different values) but the asymptotic frequency was in the range from 202.4 to 202.7 MHz.

One possible physical explanation for the unusual behavior of the calculated frequency is that in the IH-PMQ structure the electric field is concentrated mostly in the gaps, which are relatively narrow in our case. To calculate the field energy more accurately, one needs to resolve better the fields in the gaps. Unfortunately, for the hexahedral meshes it leads to a rapidly increasing mesh size.

To lower the cold model frequency when the tuner slugs are in the nominal position, the transverse cross section of the model was increased. The side panels were moved out by inserting 3-mm thick copper shims between them and the model body. The shim thickness was first estimated analytically, and then checked using the MWS computations and a frequency fit similar to the one described above. As expected, the model frequency was adjusted to 201.25 MHz.

## V. DISCUSSION

We have demonstrated that normal-conducting IH-PMQ accelerating structures are feasible and very efficient for beam velocities in the range of a few percent of the speed of light. Results of combined 3-D modeling for the IH-PMQ accelerator tank – electromagnetic computations, beam-dynamics simulations, and thermal-stress analysis – prove that H-mode structures (IH-, and the higher-mode CH-, etc) with PMQ focusing can work even at high currents. Due to the structure efficiency, the thermal management is simple and can be realized with cooling channels in the vanes. The accelerating field profile in the IH-PMQ tank was tuned to provide the best beam propagation using coupled iterations of electromagnetic and beam-dynamics modeling.

The lessons learned from the cold model fabrication and measurements help us improve the design process and will serve well for future projects. The measured field profile was in a good agreement with calculations. The deviation of the measured frequency from the calculated value led us to finding an unusual asymptotic dependence of the calculated frequency on the mesh properties, see Sec. IV.

One inherent limitation of all H-mode structures is a fixed velocity profile. Our study indicates also that some restrictions of H-PMQ structures at high currents can be caused by limited beam apertures. This is due to the fact that increasing the DT bore size forces a larger inner radius of PMQ, and increasing the inner radius of PMQ quickly weakens its focusing strength, cf. Eq. (2.1).

Within the limits of our research project we have not addressed the fabrication of the DT with embedded PMQ. After inserting the PMQ into a DT, the DT body can be closed using the electron-beam welding (EBW), with usual measures to prevent overheating the PMQ material. Similar problems have been overcome in fabrication of the SNS DTL, where the DTs contain both PMQs and cooling channels. A special iron shield was modeled [26] with the MAFIA magnetostatic solver [11] and implemented in a fixture to prevent electron-beam deviations due to the strong local magnetic fields. Of course, the DT size of the SNS DTL is larger than in our IH-PMQ structure. However, the size difference is not very large, only by a factor of 2-3, due to the DTL frequency of 402.5 MHz.

The important question of PMQ survival under long-term exposure to radiation from beam losses should be referred to the SNS DTL experience. The estimates are encouraging but the final answer will be provided by the SNS operation.

We have not addressed the structure mechanical tolerances in detail. Dedicated error studies based on multi-particle beam-dynamics simulations are required to establish the tolerances accurately.

IH-PMQ accelerating structures following a short RFQ can be used in the front end of ion linacs or in stand-alone applications. In particular, we focused on and explored a compact efficient deuteron-beam accelerator to 4 MeV. Our future plans include a study of an H-PMQ accelerator as a potential replacement for the aging LANSCE drift-tube linac. Overall, H-PMQ ion linacs look especially promising for industrial and medical applications.

## ACKNOWLEDGMENTS

The authors would like to thank James Witt for the drawings of the cold model, and Claude Conner for its assembly and alignment.

This work was supported by the US DOE via the LANL Laboratory-Directed R&D program.